\documentclass{emulateapj}

\def\arcmin{{$^{\prime}$}}
\def\arcsec{{$^{\prime\prime}$}}
\def\farcs{$''\mskip-7.6mu.\,$}

\def\Msun{\,{\rm M$_{\odot}$}}
\def\Lsun{\,{\rm L$_{\odot}$}}

\def\degr{$^{\circ}$}
\shorttitle{NGC,7538\,IRS1 - an ionized jet}
\shortauthors{Sandell et al.}

\begin{document}

\title{NGC\,7538\,IRS\,1 - an ionized jet powered by accretion}

\author{G\"oran Sandell}
\affil{SOFIA-USRA, NASA Ames Research Center, MS N211-3,
Moffett Field, CA 94035, U.S.A.}
\email{Goran.H.Sandell@nasa.gov}

\author{W. M. Goss}
\affil{National Radio Astronomy Observatory, P.O. Box O, Socorro, NM 87801, U.S.A.}
    
\author{Melvyn Wright}
\affil{Radio Astronomy Laboratory, University of California, Berkeley
601 Campbell Hall, Berkeley, CA 94720, U.S.A.}

\and

\author{Stuartt Corder\altaffilmark{1}}
\affil{Joint ALMA Observatory,  Av Apoquindo 3650 Piso 18, Las Condes, Santiago, Chile}
\altaffiltext{1}{Jansky Fellow, NRAO}

\begin{abstract}
Analysis of high spatial resolution VLA images shows that the free-free
emission from NGC\,7538\,IRS\,1 is dominated by a collimated ionized wind.
We have  re-analyzed high angular resolution VLA archive data from 6~cm
to 7~mm, and measured separately the flux density from the compact
bipolar core and the extended (1\farcs5 - 3\arcsec{}) lobes. We find
that the flux density of the core is $ \propto \nu^{\alpha}$, where
$\nu$ is the frequency and $\alpha$ is $\sim$ 0.7. The frequency
dependence of the total flux density is slightly steeper with $\alpha$ =
0.8.  A massive optically
thick hypercompact core with a steep density gradient can explain this frequency dependence, but it
cannot explain the extremely broad recombination line
velocities observed in this source. Neither can it explain why the core is bipolar rather than spherical, nor the observed 
decrease of 4\% in the flux density in less than 10 years. An ionized wind modulated by accretion
is expected to vary, because the accretion flow from
the  surrounding cloud will vary over time. BIMA and CARMA
continuum observations at 3~mm show that the free-free emission still
dominates at 3~mm. HCO$^+$ J = $1\to0$ observations combined with FCRAO
single dish data show a clear inverse P Cygni profile towards IRS\,1.
These observations confirm that IRS\,1 is heavily accreting with an
accretion rate $\sim$2 10$^{-4}$ \Msun/yr.

\end{abstract}



\section{Introduction}

NGC\,7538\,IRS\,1 was first detected in the radio at 5 GHz by
\citet{Martin73}, who found three compact radio sources at the SE edge
of the large ($\sim$ 4\arcmin{}) \ion{H}{2} region NGC\,7538, which is
at a distance 2.65 kpc \citep{Moscadelli08}. The brightest of the three,
source B, has later become known as IRS\,1 \citep{Wynn-Williams74}. The
far-infrared luminosity of the three sources is $\sim$1.9 10$^5$ \Lsun,
completely dominated by IRS\,1\citep{Hackwell82}. IRS\,1 was first
resolved with the VLA at 14.9 GHz by \citet{Campbell84}, who showed that
it has a compact ($\sim$0\farcs2) bipolar N-S core with faint extended
fan-shaped lobes, suggesting an ionized outflow. This is an extremely
well-studied source with numerous masers,  a prominent  molecular
outflow and extremely broad hydrogen recombination lines indicating
substantial mass motion of the ionized gas
\citep{Gaume95,Sewilo04,Keto08}. Although it has been modeled as an
ionized jet \citep{Reynolds86} or a photo-ionized accretion disk
\citep{Lugo04}, the source is usually referred to as an Ultra Compact or
Hyper Compact \ion{H}{2} region with a turnover at $\sim$ 100 GHz. Since
we have carried extensive observations of IRS\,1 with BIMA and CARMA,
the desire to determine the contribution from dust emission at
mm-wavelengths led us to investigate where the free-free emission from
IRS\,1 becomes optically thin. Such a determination is not possible from
published data, because IRS\,1 has been observed with a variety of VLA
configurations, some of which resolve out the extended emission and
others may only quote total flux. This prevented us from separately
estimating the flux for the compact bipolar core and the extended lobes.
Furthermore the flux density varies with time
\citep{Franco-Hernandez04}, which makes it  difficult to derive an
accurate spectral index.

\section{VLA archive data}

\begin{figure}

\includegraphics[scale=0.55,angle=-90]{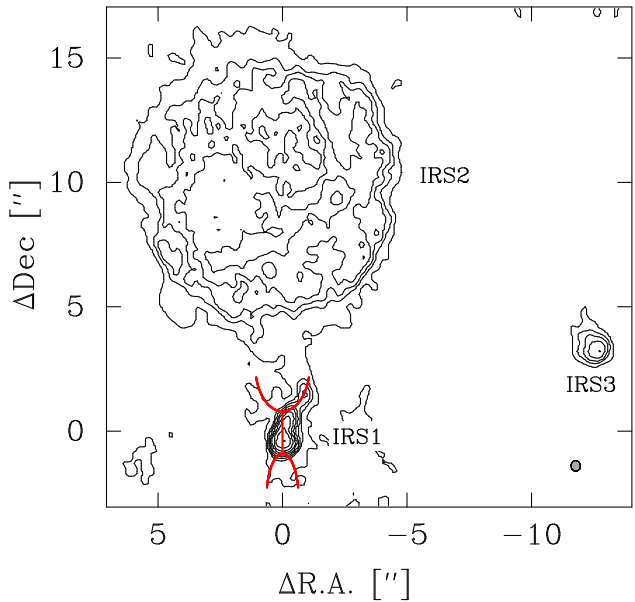}
\figcaption[]{ 
\label{fig-irs1-3} C-band (4.8 GHz)  image of the IRS1 - 3 field. The
beam size is plotted in the bottom right corner of the image. The first
contour is at 3-$\sigma$ level, the next contour 3 times higher and from
there on we plot 7 logarithmic contours. An outline of the jet from
IRS\,1, plotted in red, shows the optically thick inner part of the jet
and where the jet expands and becomes optically thin. Fig \ref{fig-vla}
zooms in on the inner optically thick part of the jet and illustrates
how the jet becomes smaller when we go up in frequency due to the outer
portion of the jet becoming optically thin.
}

\end{figure}

We have retrieved and re-analyzed a few key observations of
NGC\,7538\,IRS\,1 from the VLA data archive. These are all long
integration observations with very high angular resolution, good
uv-coverage, and high image fidelity. Table \ref{tbl-1} gives the
observing dates, synthesized beam width, sensitivity, the integrated
flux of the bipolar core, and the observed total flux. We included
a BnA array observation obtained by us at X-band (8.5
GHz)\citep{Sandell05}, because it is more contemporary with the high
frequency data sets and provides an additional data point at a
frequency where there are no other high angular resolution VLA data
available. Fig \ref{fig-irs1-3} shows the IRS\,1 - 3 field at 4.8 GHz,
while Fig \ref{fig-vla} shows the morphology of the compact bipolar core
as a function of frequency.

\section{Analysis of VLA data}
\label{Analysis}

% IRS1 position 23 13 45.371 61 28 10.42
% Kraus et al. ? = 23h13m45. s35 and ? = +61?28 10. 84 (J2000
%maser feature A (? = 23h13m45.s364, ? = +61?2810.55, J2000)
%Moscadelli 23 13 45.3622 61 28 10.507

\begin{figure}

\includegraphics[scale=0.34,angle=-90]{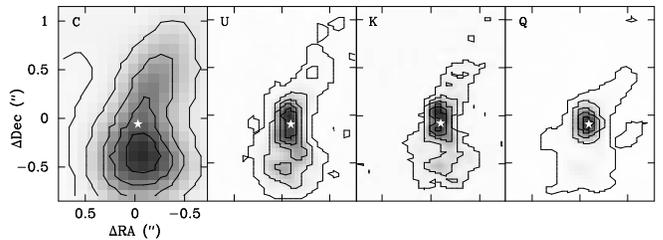}
\figcaption[]{ 
\label{fig-vla} 
VLA images at 4.86, 14.94, 22.37, \& 43.37 GHz showing the bipolar core
surrounding IRS\,1. Beam sizes are given in Table \ref{tbl-1}. The C-band image (Fig. \ref{fig-irs1-3}) shows faint extended
lobes extending several arcseconds to the north and south of IRS\,1. These images show that the core shrinks in
size with increasing frequency.}

 \end{figure}

\begin{figure}

\includegraphics[scale=0.65,angle=-90]{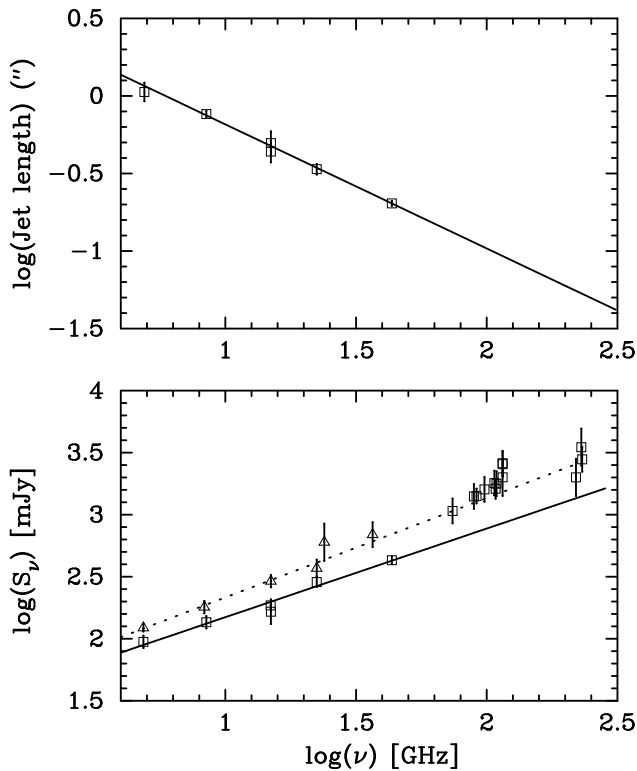}
\figcaption[]{ 
\label{fig-jet} 
{\it Top panel}: A least squares power-law fit to the length of the
optically thick jet as a function of frequency $\nu$ gives
$\theta$(arcsec) = 4.13 $\nu^{-0.8}$ where $\nu$ is in GHz. {\it Bottom
panel}: Least squares fit to VLA data  (open squares) of the compact
bipolar core surrounding IRS1 plotted as a solid line, and to total
fluxes (open triangles) plotted with a dotted line. In the figure we
also show mm data points obtained with BIMA and CARMA as well as
interferometer data from the literature, but they are not used in any of the
fits. This plot clearly demonstrates that the free-free emission still
dominates the emission in high angular resolution interferometer data
even at 1.3~mm.
}

\end{figure}

We determined flux densities of the compact bipolar core with an
accuracy of a few percent. The results are given in Table \ref{tbl-1}.
Table \ref{tbl-1} also gives total flux densities, obtained by
integrating over the whole area where we detect emission from IRS\,1. At
4.9 GHz the total linear extent of IRS\,1 is $\sim$ 7\arcsec, while it
is $\sim$  1\farcs4 at 43.4 GHz. Even though we can reliably determine
the flux density of the bipolar core at 43.4 GHz, some of the faint
extended emission is filtered out; therefore the total flux is
underestimated. At 4.9 GHz the bipolar core has a total length of $\sim$
1\farcs1 FWHM (Full Width Half Maximum), with a separation between the
two peaks of  0\farcs68, at 14.9 GHz the size is 0\farcs44 with a peak
separation of  0\farcs24 (Table \ref{tbl-1}). At 43.4 GHz we can still
fit the core with a double Gaussian with a peak separation of $\sim$
0\farcs12, while the linear size (length) has a FWHM of  0\farcs20.
Since one of our data sets ( X-band, 8.4 GHz) has  insufficient
resolution to resolve the two bipolar cores, we plot the linear size
(length) as a function of wavelength in Fig. \ref{fig-jet}, although fitting to the lobe separation gives
virtually identical results, but with a larger uncertainty. We find the size to vary with frequency as $\nu^{-0.8 \pm 0.03}$, while the flux
density increases with frequency as $\nu^{0.7 \pm 0.05}$. The frequency
dependence of the total flux is more difficult to estimate, since some
VLA configurations are not very sensitive to faint extended emission and
will therefore underestimate the total flux. If we chose data sets which
give a reliable flux for IRS\,2\footnote{The emission from  IRS\,2 is
already optically thin at 4.8 GHz (C-band) with an integrated flux of
1440 mJy.}, which is a spherical \ion{H}{2} region with a size of $\sim$
8\arcsec, we find that the total flux from IRS\,1 has a slightly steeper
spectral index,  0.8 $\pm$ 0.03 determined from data sets from 4.8 GHz to 49
GHz. The fit to the total flux densities for IRS\,1 is shown as a dotted
line in Fig. \ref{fig-jet}. Such a shallow frequency dependence would
require a very steep density gradient in the ionized gas, if the
emission originates from an \ion{H}{2} region ionized by a central
O-star, although it is not impossible, see e.g. \citet{Keto08}. A high-density
\ion{H}{2} region with a steep density gradient should be spherical, not bipolar with a dark central lane as we observe in IRS\,1. Neither can a steep density gradient model explain the extremely broad recombination
lines observed in IRS\,1 \citep{Gaume95, Keto08}. All these characteristics
can be explained,  if the emission originates in an ionized stellar wind or jet \citep{Reynolds86}. The inner part of the jet is optically
thick and further out, where the jet expands, the emission becomes
optically thin. At higher frequencies the outer part of the jet becomes
optically thin and therefore appears shorter and more collimated,
exactly what we see in IRS\,1, see Fig \ref{fig-vla}. For a uniformly
expanding spherical wind, the size of the source varies as a function of
frequency as $\nu^{-0.7}$, while the flux density density goes as
$\nu^{0.6}$ \citep{Panagia75}. Since we resolve the emission from
IRS\,1, we know that it is not spherical, but instead it appears to
originate in an initially collimated bipolar jet (opening angle $\lesssim$ 30\degr{})
approximately aligned with the molecular outflow from IRS\,1, see Section~\ref{accretion}. For a
collimated ionized jet, \citet{Reynolds86} showed that it can have a
spectral index anywhere between 2 and $-$0.1, depending on gradients in
jet width, velocity, degree of ionization, and temperature.
% For a
%collimated, pressure confined jet,  which is very plausible for IRS\,1,
%his model predicts a spectral index of 0.84, which is very close to what
%we find.

\begin{deluxetable*}{lccllll}
\tabletypesize{\scriptsize}
\tablecolumns{7}
\tablenum{1}
\tablewidth{0pt} 
\tablecaption{Observational parameters and results of VLA archive data of IRS\,1 \label{tbl-1}} 

\tablehead{
\colhead{Frequency}  & \colhead{Synthesized beam} & \colhead{rms}& \colhead{Size} & \colhead{flux density}& \colhead{flux density} & \colhead{Observing dates}\\
\colhead{}                      & \colhead{}                                  & \colhead{}        &\colhead{bipolar core}  & \colhead {bipolar core} & \colhead{total}            & \\
\colhead{[GHz]} & \colhead{\arcsec{} $\times$ \arcsec{} pa = \degr{} }   &\colhead{[mJy~beam$^{-1}$]} & \colhead{\arcsec{} $\times$ \arcsec{} pa = \degr{} } & \colhead{[mJy]} & \colhead{[mJy]} & \colhead{}
}
\startdata
\phantom{0}4.86 &  0.43 $\times$ 0.37 ~~$+$\phantom{0}5.4\degr\   & 0.30\phantom{0} & 1.06 $\times$ 0.44 ~~$-$9.6 & \phantom{0}94.3 $\pm$ 4.8  &  \phantom{00}122.5 & 1984:1123 \\
%\phantom{0}4.89  & 1.35 $\times$ 1.09 ~~$-$24.0\degr\ & 0.10\phantom{0} &  \phantom{0}89.8  $\pm$ 9.0\tablenotemark{a}   &  \nodata\tablenotemark{a}       & 19890428\\
\phantom{0}8.46 &  1.21 $\times$ 0.47 ~~$-$\phantom{0}0.5\degr\           &  0.05\phantom{0} & 0.76 $\times$ 0.30 ~~$-$10.9 &135.7   $\pm$ 13.6\tablenotemark{a} &  \phantom{00}148.1\tablenotemark{a}  & 2003:1014\\
14.94                    &  0.14 $\times$ 0.12 ~~$+$47.6\degr\               &  0.017  & 0.44 $\times$ 0.20 ~~$-$0.3 & 185.9 $\pm$  5.6 &  \phantom{00}248.7 & 2006:0226,0512\tablenotemark{b} \\
14.91                    &  0.14 $\times$ 0.10 ~~$+$13.7\degr\    & 0.081 & 0.50 $\times$ 0.20 ~~$-$1.2 &165.2  $\pm$ 13.3 &  \phantom{00}291.0 & 1994:1123\\
22.37                    & 0.08 $\times$ 0.08 ~~$-$43.0\degr\      & 0.017 &  0.34 $\times$ 0.17 ~~$+$3.1 & 286.5 $\pm$ 11.5 &  \phantom{00}370.0 & 1994:1123\\
43.37                    & 0.14 $\times$ 0.12 ~~$+$14.5\degr\    & 0.17\phantom{0} & 0.20 $\times$ 0.14 ~~$-$1.0 & 429.7 $\pm$ 12.9 &  \phantom{00}473.7   & 2006:0706,0914\tablenotemark{b} \\
\enddata

\tablenotetext{a} {Insufficient spatial resolution, blends in with IRS\,2. Unreliable  total flux.}
\tablenotetext{b} {Average of two data-sets at different epochs during the same year.}
\end{deluxetable*}

\section{Outflow and accretion}
\label{accretion}

\begin{figure}

\includegraphics[scale=0.45]{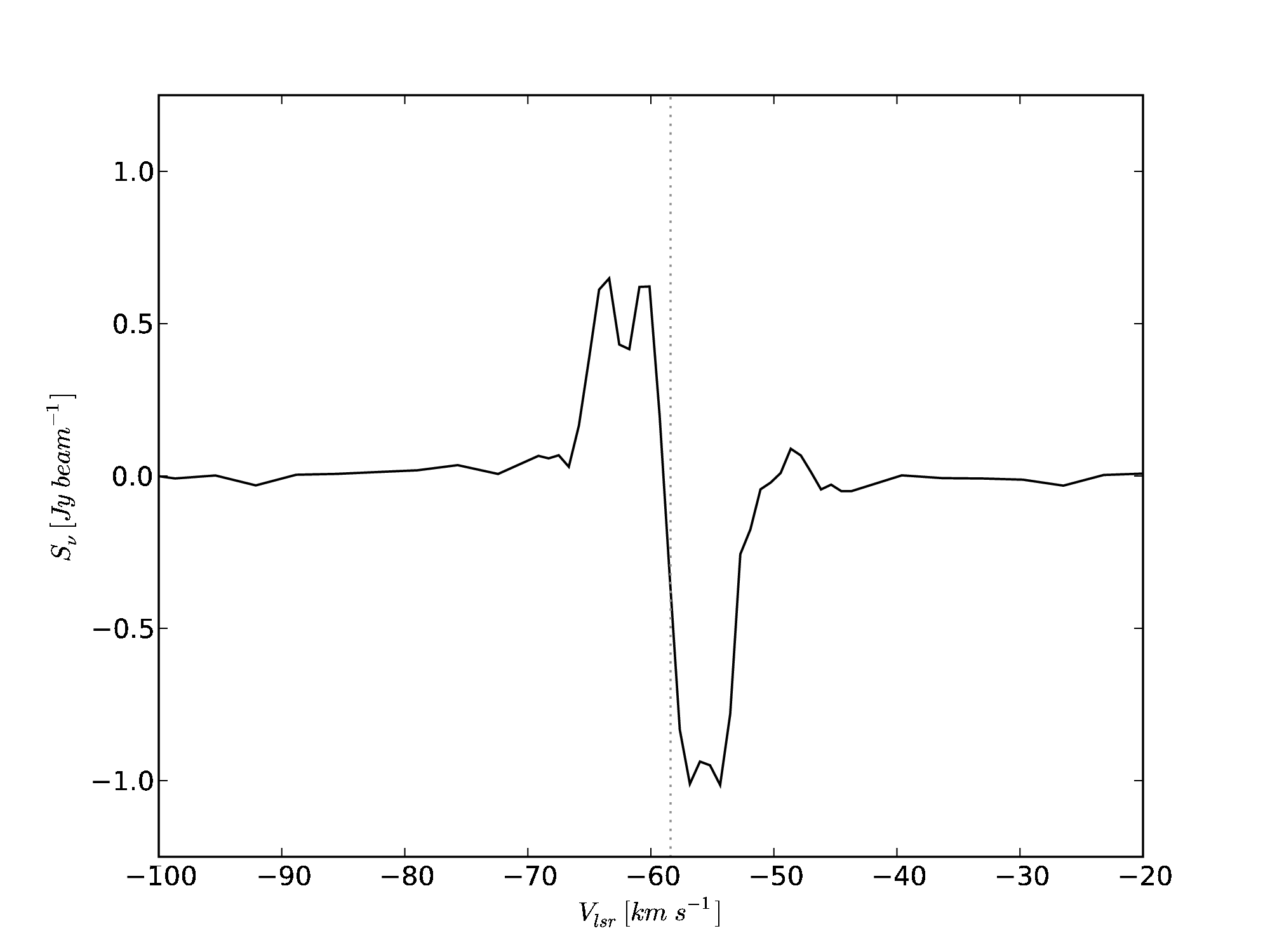} 
\figcaption[]{
\label{fig-IPC} 
Inverse P Cygni profile in HCO$^+$ J = $1 \to 0$ observed with CARMA
supplemented with FCRAO single dish data \citep{Corder08}. Note that the
spectrum is extracted from a continuum subtracted spectral line cube,
hence the flux density appears to go below zero. The angular resolution
is 4\farcs5. The vertical dotted line shows the systemic velocity of IRS\,1.
}
\end{figure}

Although the molecular outflow from IRS\,1 in early studies was found to
be quite compact and going from SE to NW \citep{Scoville86}, we find
that the molecular outflow is very extended  ($>$ 4\arcmin{}) and
approximately aligned with the free-free jet from analyzing large
mosaics obtained with CARMA in  $^{12}$CO and $^{13}$CO J = $1 \to 0$,
and HCO$^+$ J = $1 \to 0$. For HCO$^+$ we filled in missing zero spacing
with fully sampled single dish maps obtained with FCRAO. A thorough
discussion of the molecular outflow is beyond the scope of this paper,
and will be discussed in a forthcoming paper (Corder et al. 2009, in
prep). Here we therefore summarize some of the main results. The CARMA
mosaics show that the outflow is very large and has a position angle
(pa) of  $\sim$ $-$20\degr,  which is similar to the orientation of the
collimated inner part of the ionized jet driven by IRS\,1 on smaller
angular scales. The outflow extends several arcminutes to the north and
starts as a wide angle limb brightened flow. To the south the outflow is
difficult to trace, because of several other outflows in the giant
molecular core in which IRS\,1 is embedded, but appears more collimated
with a linear extent of $\sim$ 2\arcmin. Analysis of Spitzer IRAC
archive data show that the outflow may be even larger to the south. The
5.8 and 8.0 $\mu$m IRAC images show a jet-like feature projecting back
towards IRS\,1,  extending to $\sim$ 3.5\arcmin\ (2.6 pc) from IRS\,1,
which is much further out than what was covered by the CARMA mosaics
\citep{Corder08}. However, the CARMA observations also confirm that
IRS\,1 is still heavily accreting. In HCO$^+$ J = $1 \to 0$ we observe a
clear inverse P Cygni profile towards the strong continuum emission from
IRS\,1 (Fig \ref{fig-IPC}).  The absorption is all red-shifted and has
two velocity components, which could mean that the accretion activity is
varying with time (episodic). From these data \citet{Corder08} estimates
$\sim$ 6.8 \Msun\ of infalling gas. With the assumption that the accretion
time is similar to the free-fall time, $\sim$ 30,000 yr,
\citeauthor{Corder08} finds an accretion rate $\sim$ 2 10$^{-4}$ \rm
M$_{\odot}$/yr,  which will block most of the uv photons from the central
O-star, allowing them to escape only at the polar regions.

\section{Millimeter data - where is the accretion disk? }

In Fig. \ref{fig-jet} we also plotted flux densities at 3 and 1~mm from
the literature \citep{Lugo04}, supplemented with our own results at 3~mm
from BIMA (Wright et al. 2009, in prep) and CARMA \citep{Corder08}. All
the observed flux densities at 3~mm or even at 1.3~mm can be explained
by free-free emission, with at most a marginal excess from dust
emission. High angular resolution CARMA continuum observations at 91.4
and 108.1 GHz confirm what is predicted from the fit to total flux
densities in Fig~\ref{fig-jet}, i.e. that the free-free emission
dominates at 3~mm. These CARMA observations resolve IRS\,1 with a size
of 0\farcs5 $\times\leq$0\farcs3 pa $-$1\degr\ and 0\farcs4
$\times\leq$0\farcs1 pa 20\degr\ at 94 and 109 GHz, respectively, i.e.
the 3~mm emission is aligned with the free-free emission, and not with
an accretion disk, which is expected to be perpendicular to the jet.
Since  we detect a collimated free-free jet and since there is a strong
accretion flow towards IRS\,1 (Section \ref{accretion}), IRS\,1 must be
surrounded by an accretion disk. The morphology of the free-free
emission suggests that the disk  should be almost edge-on, and centered
halfway between the northern and the southern peak of the optically
thick inner part of the free-free jet. However, if such an edge-on
accretion disk is thin, i.e. the height of the disk is small relative to
its diameter,  it may not provide much surface area, and is therefore
difficult to detect at 3~mm.  At 3~mm IRS\,1 and
IRS\,2 are very strong in the free-free, and even the free-free emission
from IRS\,3 cannot be ignored, which makes it very hard to detect dust
emission from the accretion disk. Since the spectral index for dust is
$\sim$ 3 - 3.5, the dust emission may start to dominate at frequencies
above 300 GHz, observable only by SMA, but even at 1.3~mm we may have a
much better chance to detect the accretion disk. There is some evidence
for such a disk in 450 and 350 $\mu$m continuum images obtained at JCMT
\citep{Sandell04}. Until higher spatial resolution interferometer images
are available, it is not clear whether this emission originates in a
disk, from the surrounding in-falling envelope, or from a superposition
of several nearby sources.

 \section{Discussion and Conclusions}
 
Free-free emission from ionized jets is very common in low mass
protostars (Class 0 and Class I sources), which all appear to drive
molecular outflows. This situation is especially true for young stars, which often have rather well
collimated outflows \citep{Rodriguez99}. Jets may also be
common  in young early B-type high mass stars \citep{Gibb07}, although
they have not been studied as well as low-mass protostars, because they
are  short lived, more distant and harder to identify. Although
there have not been any detections  of jets in young O stars, i.e. stars
with a luminosity of $>$ 10$^5$ \Lsun, such objects are likely to exist.
Broad radio recombination line (RRL) objects, many of which are
classified as Ultra Compact or Hyper Compact \ion{H}{2} regions
\citep{Jaffe99,Kurtz02,Sewilo04,Keto08} offer a logical starting point, because
they show evidence for substantial mass motions, which would be readily
explained if the recombination line emission originates in a jet.
Several of them also show evidence for accretion. In the sample
analyzed by \citet{Jaffe99}, which partly overlaps with the sources
discussed by \citet{Gibb07}, they find four sources (including IRS\,1)
with  bipolar morphology. All appear to be dominated by wind ionized
emission, although not necessarily jet driven winds. K\,3-50\,A,
however,  at a distance of  8.7 kpc \citep{DePree94}, has a luminosity
of a mid-O star, drives an ionized bipolar outflow \citep{DePree94}, and
has a spectral index of  0.5 in the the frequency range 5 - 15 GHz
\citep{Jaffe99}. K\,3-50\,A is therefore another example of an O-star,
where the free-free emission appears to be coming from an ionized jet.
Detailed studies of broad RRL objects will undoubtedly discover more
examples of jet-ionized \ion{H}{2} regions. Such objects are likely to
drive molecular outflows, excite masers and show evidence for strong
accretion.
 
To summarize: We have shown that the emission from IRS\,1 is completely
dominated by a collimated ionized jet. The jet scenario also readily explains why the free-free
emission is variable, because the accretion rate from the surrounding clumpy molecular cloud will vary with time. It also explains why IRS\,1 is an extreme broad RRL object. Since most
Hyper Compact \ion{H}{2} regions are broad RRL objects, and show rising
free-free emission with similar spectral index to IRS\,1, other sources
now classified as Hyper Compact \ion{H}{2} regions, may also be similar to IRS\,1.

\acknowledgements The National Radio Astronomy Observatory (NRAO)  is a facility of the National Science
Foundation operated under cooperative agreement by Associated
Universities, Inc. The BIMA array was operated by the Universities of
California (Berkeley), Illinois, and Maryland with support from the
National Science Foundation. Support for CARMA construction was derived
from the states of California, Illinois, and Maryland, the Gordon and
Betty Moore Foundation, the Kenneth T. and Eileen L. Norris Foundation,
the Associates of the California Institute of Technology, and the
National Science Foundation. Ongoing CARMA development and operations
are supported by the National Science Foundation under a cooperative
agreement, and by the CARMA partner universities.

\end{document}